# SUPER-NYQUIST CO-PRIME SENSING

*Abstract*
*The theory of co-prime arrays has been studied in the past. Nyquist rate estimation of second order statistics using the combined difference set was demonstrated with low latency. This paper proposes a novel method to reconstruct the second order statistics at a rate that is twice the Nyquist rate using the same sub-Nyquist co-prime samplers. We analyse the difference set, and derive the closed-form expressions for the weight function and the bias of the correlogram estimate. The main lobe width of the bias window is approximately half of the width obtained using the prototype co-prime sampler. Since the proposed scheme employs the same rate prototype co-prime samplers; the number of samples acquired in one co-prime period and hardware cost are unaffected. Super-Nyquist estimation with multiple co-prime periods is also described. Furthermore, n-tuple or multi-level co-prime structure is presented from a super-Nyquist perspective. Here, estimation at a rate q times higher than Nyquist is possible, where q is the number of sub-samplers.*

## 1. INTRODUCTION

Analog signal acquisition is a well-studied topic. A detailed survey of the sampling techniques for acquisition can be found in [1]. It also describes sub-Nyquist strategies. Initial work on Nyquist rate acquisition is discussed in [2]-[5]. Nyquist rate acquisition requires the sampling frequency $f_s$ to be at least twice the bandwidth of the signal. Practically, $f_s$ is often chosen to be greater than twice the bandwidth.

The nested arrays [6] and co-prime arrays [7] were proposed as efficient sub-Nyquist strategies for estimation of second order statistics. Prior to this, the minimum redundancy arrays were also considered [8]. This paper will focus on the co-prime structure and propose a modification. The prototype co-prime sampling structure has two samplers operating at a rate M and N times lower than the Nyquist rate. (M, N) is a co-prime pair. Each sampler has a uniform sampling structure. It can therefore be implemented using traditional low rate analog-to-digital converters. Low latency estimation was discussed in [9]-[11]. Spectral estimation is well studied for the Nyquist case [12][13], and has also been investigated for co-prime array using the correlogram method with low latency [14]. This will form the basis for the spectral theory and derivation presented in this paper.

Several modified co-prime structures have been proposed and estimation aspects discussed, for example, extended co-prime arrays [15]-[17], Co-prime array with compressed inter-element spacing (CACIS), the co-prime array with displaced sub-arrays (CADiS) [18], thinned co-prime arrays [19], co-prime arrays with multiple periods [20],etc. However, they may not appear to be an attractive option for temporal sampling since their structure does not provide uniformity in sampling patterns for the individual samplers. In other words, one of the samplers will need to be turned off for a certain duration in each co-prime period. Nevertheless, all these structures can be considered for super-Nyquist implementation. However, this paper will introduce the concept of super-Nyquist estimation for structures which provide uniformity in sampling patterns. A summary is provided below:

- This paper proposes a novel super-Nyquist co-prime sampling scheme to reconstruct the second-order statistics at twice the Nyquist rate using the same sub-Nyquist samplers employed by the prototype co-prime scheme.

- In other words, if the prototype co-prime sampler can estimate the second order statistics of a signal with bandwidth B, then the super-Nyquist scheme can estimate for signals with bandwidth 2B.

- The weight function and associated correlogram bias window expressions are derived.

- The proposed scheme can be related to the work in [21] and will be discussed briefly.

- The super-Nyquist co-prime samplers with multiple periods is also described.

Author: Usham V. Dias
Dept. of Electrical Engineering, Indian Institute of Technology Delhi.

- Furthermore, super-Nyquist scheme with multiple sub-samplers is described. This structure is capable of estimating the statistics of a signal with bandwidth qB. Here, q is the number of sub-samplers.

- A discussion on future research challenges is also provided.

Section 2 describes the novel super-Nyquist structure, Section 3 analyzes the difference set, Section 4 derives the bias of the correlogram method, Section 5 presents the simulation results, and Section 6 discusses the multiple period scenario with multiple sub-samplers. This is followed by concluding remarks and future directions in Section 7.

## 2. STRUCTURE

The sampling pattern for the prototype or traditional co-prime samplers is shown in Fig. 1(a), where (*M*, *N*) represents the co-prime pair and *d* the Nyquist sampling period. The combined samples from the individual co-prime samplers do not contain all the possible Nyquist samples. However, despite the missing samples, the structure has the ability to reconstruct the autocorrelation at the Nyquist rate. This is well studied in the literature [7][14]. The number of samples acquired in one co-prime period *MNd*, is *M+N*. It may be noted that the traditional co-prime sampling structure has the zeroth sample coinciding. This may seem to be wasteful since both the samplers acquire the signal at the same instant of time.

The novel super-Nyquist co-prime sampler that is proposed here, is designed such that the zeroth sample does not coincide. This is shown in Fig. 1(b). The value of (*M*, *N*) is taken as (4, 3) to explain the concept. Note that the second sampler is placed at an offset of $\frac{d}{2}$, where $\frac{d}{2}$ is the super-Nyquist sampling time period. The idea is to place the second sub-sampler at half the Nyquist sampling period. Thus, the new sampling rate is twice the Nyquist rate i.e. $2f_s$. In a single co-prime period, the sampling times or instants at which the first sampler acquires the signal is given by, *Mnd*, where n ϵ [0, N-1], i.e. [0, Md, 2Md, ... (N-1)Md]. The second sampler acquires samples at times $Nmd+\frac{d}{2}$, where m ϵ[0, M-1], i.e. [0.5d, (N+0.5)d, (2N+0.5)d, ... ((M-1)N+0.5)d].

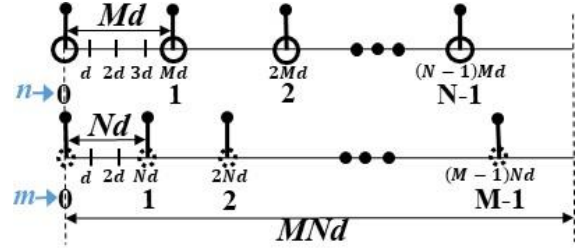

(a) Traditional co-prime sampling

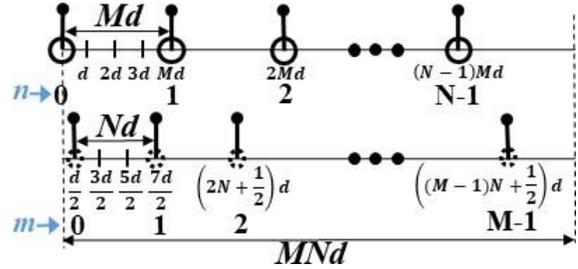

(b) Super-Nyquist co-prime sampling

Fig. 1: Traditional and super-Nyquist co-prime structures.

As an example, if *M=4*, *N=3* and *d=1*, the acquired sample locations are [0, 4, 8] and [0.5, 3.5, 6.5, 9.5] for the first and second sub-sampler respectively.

Note that the second sampler can be kept fixed, while offsetting the first by $\frac{d}{2}$. As long as the difference between the first sample of each individual samplers is $\pm\frac{d}{2}$, super-Nyquist reconstruction of second order statistics is possible.

This structure can be viewed as a special case of the co-prime sampler with jitters in the sampling location [21], if the jitters suffered by the first and second samplers are constants (i.e. $\epsilon_1$ and $\epsilon_2$) and the difference between them is $\pm\frac{d}{2}$ (i.e. $\epsilon_1 - \epsilon_2 = \pm\frac{d}{2}$). It implies that the two samplers are non-synchronized with an offset of $\pm\frac{d}{2}$.

## 3. DIFFERENCE SET

This section describes the difference set for the proposed super-Nyquist scheme. The self differences for the super-Nyquist scheme is same as that of the prototype co-prime array. $\mathcal{L}^+{}_{SM}$ represents the self difference set for the sampler $x(Mn)$ and $\mathcal{L}^{+}_{SN}$ for the sampler $x(Nm+0.5)$. $\mathcal{L}^{-}_{SM}$ and $\mathcal{L}^{-}_{SN}$ represent their negative sets respectively.

$$\mathcal{L}^{+}_{SM} \cup \mathcal{L}^{-}_{SM} = Mn_1 - Mn_2 \ \{n_1, n_2 \in [0, N-1]\}$$

$$\mathcal{L}^{+}_{SN} \cup \mathcal{L}^{-}_{SN} = (Nm_1 + 0.5) - (Nm_2 + 0.5) \ \{m_1, m_2 \in [0, M-1]\} \quad (1)$$

The self difference matrices $\mathcal{L}^{+}_{SM} \cup \mathcal{L}^{-}_{SM}$ and $\mathcal{L}^{+}_{SN} \cup \mathcal{L}^{-}_{SN}$, as shown in Fig. 2(a)-2(b), are same as that of the prototype co-prime scheme. The number of sample pairs available to estimate the second-order statistics at each self difference value is given by the number of times a self difference repeats in the matrix, denoted within the braces {•}. The cross difference is given by:

$$\mathcal{L}^{+}_{C} = Mn - (Nm + 0.5) \text{ and } \mathcal{L}^{-}_{C} = (Nm + 0.5) - Mn \quad (2)$$

The cross difference set $\mathcal{L}^{+}_{C}$ is shown in Fig. 2(c)-(f) for four different values of (M, N). It includes two cases each with M > N and N > M. The set $\mathcal{L}^{-}_{C}$ is the negative of the values in $\mathcal{L}^{+}_{C}$.

**Claim 1:** The self differences do not form a subset of the cross differences for the super-Nyquist scheme, i.e. $\mathcal{L}_S \nsubseteq \mathcal{L}_C$.

*Proof:* Let $l_c = Mn - (Nm + 0.5)$ be an element in the super-Nyquist cross difference set $\mathcal{L}^{+}_{C}$. Substituting m=0 in this equation gives $l_c = Mn - 0.5 \neq Mn$ and hence, does not belong to $\mathcal{L}^{+}_{SM}$. Substituting n=0 gives $l_c = -Nm - 0.5 \neq$ -Nm and hence, does not belong to $\mathcal{L}^{-}_{SN}$. Furthermore, substituting any value of m ϵ [0, M-1] and n ϵ [0, N-1] does not generate an integer because of the constant '0.5' in the equation for $l_c$. Thus, self differences, which are integers, can never be generated. A similar argument can be made for the set $\mathcal{L}^{-}_{C}$, thus proving the claim.

**Claim 2:** The cross difference set $\mathcal{L}^{+}_{C}$ (and $\mathcal{L}^{-}_{C}$) has MN distinct values.

Claim 2 can be easily proved along similar lines as shown for the prototype co-prime arrays. In addition to the above claim, it is worth noting that the cross difference set does not have a uniform behavior. For the case when (M, N) = (4,3), the set $\mathcal{L}^{+}_{C}$ has three difference values appearing along with its negative value in the same set, viz. (0.5, -0.5), (1.5, -1.5) and (3.5, -3.5) as shown in Fig. 2(c), while the remaining elements do not have a negative pair. The paired elements have two contributors for autocorrelation estimation at these difference values, while the

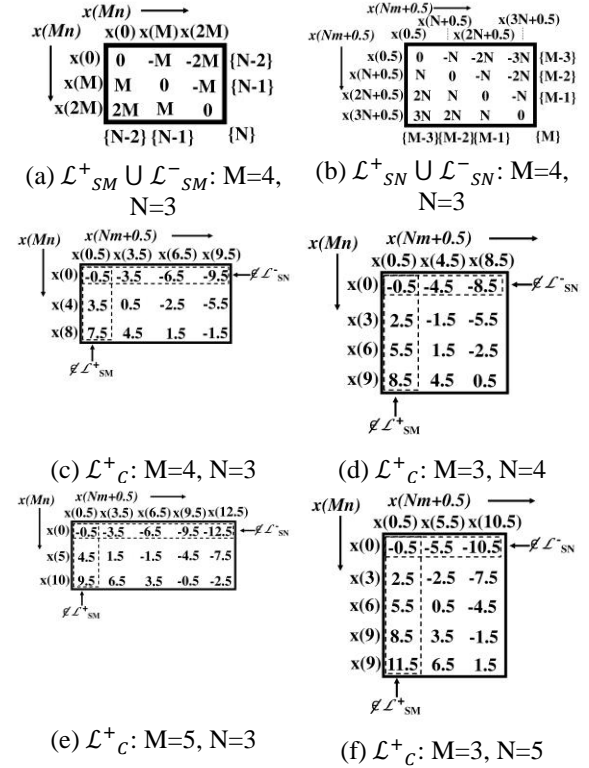

(a) $\mathcal{L}^{+}_{SM} \cup \mathcal{L}^{-}_{SM}$: M=4, N=3
(b) $\mathcal{L}^{+}_{SN} \cup \mathcal{L}^{-}_{SN}$: M=4, N=3
(c) $\mathcal{L}^{+}_{C}$: M=4, N=3
(d) $\mathcal{L}^{+}_{C}$: M=3, N=4
(e) $\mathcal{L}^{+}_{C}$: M=5, N=3
(f) $\mathcal{L}^{+}_{C}$: M=3, N=5

Fig. 2: Super-Nyquist co-prime samplers: Difference sets.

unpaired have only one contributor. For the cases when (M, N) takes values (5,3), (3,4) and (3,5), there are six, six and four pairs respectively in the set $\mathcal{L}^{+}_{C}$ as shown in Fig. 2(d)-(f).

In contrast, the prototype co-prime sampler has a cross difference set with a more uniform behaviour since there are exactly two contributors for autocorrelation estimation at each difference value in set $\mathcal{L}_C - \mathcal{L}_S$ (refer Proposition III in [11]).

The number of contributors for autocorrelation estimation at difference value '$l$' (denoted by $z(l)$) is shown in Fig. 3 for the prototype co-prime sampler and in Fig. 4 for the super-Nyquist co-prime sampler. The difference values in the super-Nyquist set are non-integers and have been mapped to integer values based on the following map:

$$l = f(l_r) = 2l_r \text{ where } f: \mathbb{R} \mapsto \mathbb{Z} \quad (3)$$

where $f$ is a function that maps the real numbers $l_r$ in the super-Nyquist difference set to integer values $l$. This function is a one to one map. Therefore, each integer, $l$, represents a lag which is a multiple of $\frac{d}{2}$ for the super-Nyquist scheme. Interchanging the values of M and N does not change the function $z(l)$ for the prototype co-prime array, however, it affects the super-

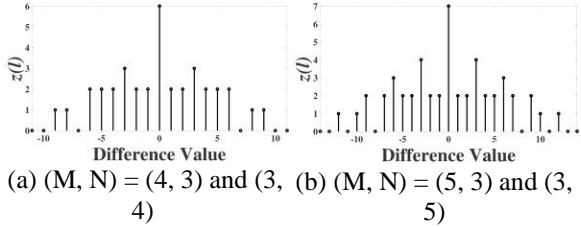

(a) (M, N) = (4, 3) and (3, 4)  (b) (M, N) = (5, 3) and (3, 5)

Fig. 3: Weight function for prototype co-prime sampler.

Nyquist scheme. This effect is due to the fact that the second array with inter-element spacing of Nd has an

Offset. Therefore, a change in N causes a change in the locations at which the actual samples are acquired, which is not the case for the prototype co-prime array (refer Fig. 1).

It is evident that this scheme is capable of generating difference values at twice the Nyquist rate but has some missing difference values. Such missing values were also observed in the prototype co-prime array and its generalizations. Note that the nested array is the only structure that generates a filled difference set [6]. However, the work in [14] has shown that the entire difference set including holes (missing values) can be used to estimate the spectral content of a signal. It also guarantees the positiveness of the estimate. In addition, there are researchers investigating methods to mitigate the effect of holes. Some of the techniques are nuclear norm minimization method [22], positive definite Toeplitz completion [23], array interpolation [24] and convex sparse recovery [25].

## 4. CORRELOGRAM BIAS WINDOW

Missing autocorrelation values may be a source of error in the estimation process. However, for correlogram spectral estimation, the shape of the function $z(l)$ affects the fidelity of the spectral estimate. The Fourier transform of $z(l)$ represents the correlogram bias window. The correlogram theory for the Nyquist case is presented in [12][13]. It has been recently analyzed for the co-prime sampler/arrays in [14]. It was shown that the sub-Nyquist correlogram estimate can be approximated by the convolution of the Nyquist spectrum with the bias window. Therefore, we would ideally want the bias window to be an impulse function. Practically, this is not possible.

The closed-form expression of $z(l)$ for the proposed super-Nyquist scheme can be derived from claims 1-

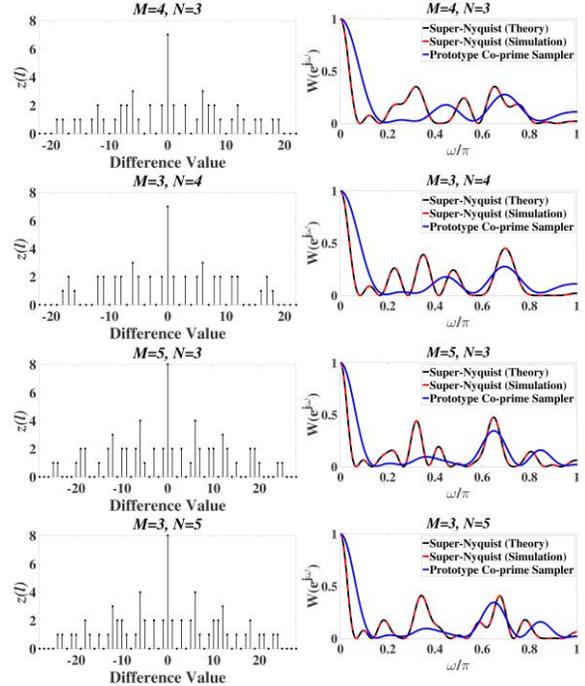

Fig. 4: Super-Nyquist co-prime sampler: Weight function and normalized bias curves.

2, and is given by (4). The correlogram bias window is given by (5). Refer Appendix A for the derivation.

This theoretical bias equation is verified with the simulated bias (FFT of the weight function). It is also compared with the bias of the prototype co-prime sampler in Fig. 4. Since the bias is symmetric, only one half is shown in the normalized range [0, 1]. Note that the bias for the prototype co-prime sampler is the same when (M, N) is equal to (4, 3) and (3, 4) (also for (5, 3) and (3, 5)). This is as expected since the function $z(l)$ is the same when the values of M and N are interchanged for the prototype co-prime sampler. However, it does not hold true for the super-Nyquist scheme.

For all the cases considered in Fig.4, it is evident that the main lobe width of the super-Nyquist bias window is approximately one half of the prototype co-prime width. This implies better resolution. Furthermore, there is no additional acquisition cost when compared to prototype co-prime scheme. The same rate analog-to-digital converters are employed with the same number of acquired samples in one co-prime period, i.e. M+N. However, some additional side-lobes are observed in the super-Nyquist scheme and will be shown to have no serious consequences on spectral peak estimation. In addition, as mentioned earlier, the research on autocorrelation estimation at the missing difference values or holes could have a positive effect on side-lobe reduction. The use of an appropriate window function can also serve the purpose.

$$z(l) = \sum_{n=-N+1}^{N-1}(N-|n|)\delta(l-2Mn)$$
$$+ \sum_{m=-M+1}^{M-1}(M-|m|)\delta(l-2Nm)$$
$$+ \sum_{n=0}^{N-1}\sum_{m=0}^{M-1}\delta(|l|-|2Mn-2Nm-1|)$$

(4)

$$W(e^{j\omega}) = \frac{1}{s}\left\{\left|\frac{\sin\omega MN}{\sin\omega M}\right|^2 + \left|\frac{\sin\omega MN}{\sin\omega N}\right|^2 + 2\cos(\omega(M-N+1))\frac{(\sin\omega MN)^2}{\sin\omega M \sin\omega N}\right\}$$

(5)

## 5. SIMULATION RESULTS

Before discussing the results, let us try to understand the signal model and the simulation setup. The focus here is on temporal signal acquisition and correlogram spectral estimation. The signal model employed is similar to that described in [14][26].

Let us assume that the input signal is made up of frequency bands at 50Hz and 150Hz. Let us assume the sampling frequency $f_s = 500$Hz. Therefore, the Nyquist sampling period is $d = \frac{1}{f_s}$. Here, the maximum frequency that can be measured is $f_{max} = \frac{f_s}{2} = 250$Hz. This is the Nyquist framework. For the sub-Nyquist prototype co-prime scheme, as noted in Fig. 1(a), we need to skip the Nyquist samples based on factors (M, N) which represents the actual sampling. Here, $f_s$ can be viewed as the virtual sampling rate.

However, the proposed super-Nyquist scheme acquires samples at integer multiples of d as well as at fractions as described in Fig 1(b). Note that the smallest distance between two samples is $d_{ss} = 0.5d$. This implies that the super-Nyquist virtual sampling frequency is $f_{ss} = \frac{1}{d_{ss}} = 2f_s$. The actual sampling rate is same as the prototype co-prime. Therefore, when $f_s = 500$Hz, we have $f_{ss} = 1000$Hz. Now, the highest frequency that can be measured is $f_{max_{ss}} = \frac{f_{ss}}{2} = 500$Hz.

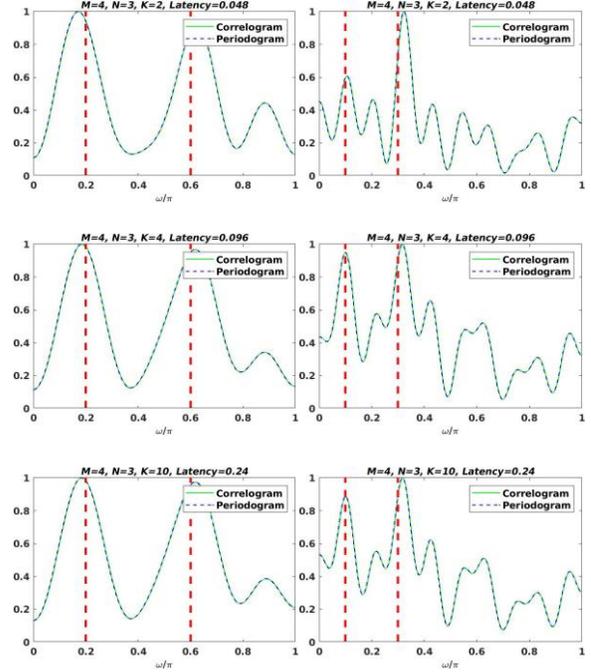

Fig. 5: Spectral estimation using super-Nyquist and prototype co-prime samplers.

Table 1: Relationship between actual and normalized frequencies.

| Hertz | | 50 | 150 | 250 | 300 | 450 | 500 |
|---|---|---|---|---|---|---|---|
| **Normalized** | Super-Nyquist | 0.1 | 0.3 | 0.5 | 0.6 | 0.9 | 1 |
| | Prototype | 0.2 | 0.6 | 1 | - | - | - |

As we know, the spectrum of a discrete signal repeats with a period $2\pi$. Furthermore, for real signals the spectrum is symmetric, i.e. the contents in the range $(0, \pi)$ is also available in range $(\pi, 2\pi)$. Therefore, the range $(0, \pi)$ is sufficient to describe the results. In terms of frequency in Hertz and normalized frequency, this corresponds to $(0, \pi) = (0, f_{max} = \frac{f_s}{2}) = (0, 1)$ for the Nyquist and prototype co-prime scheme. It corresponds to $(0, \pi) = (0, f_{max_{ss}} = \frac{f_{ss}}{2}) = (0, 1)$ for the super-Nyquist scheme.

Here, the normalized frequency is used for the simulation plots. Therefore, the frequencies at 50Hz and 150Hz will appear at different locations in the normalized frequency plot for the prototype co-prime and super-Nyquist co-prime scheme. Please refer Table 1 for the relationship between actual and normalized frequencies.

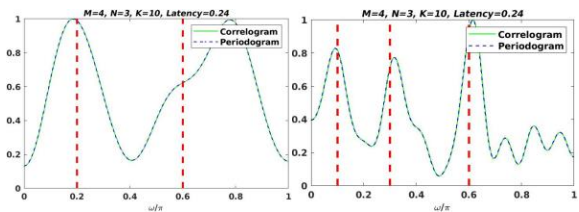

Fig. 6: Spectral estimation using super-Nyquist and prototype co-prime samplers: Three peaks, (M=4, N=3).

The results in Fig.5 for spectral estimation indicate that the prototype and super-Nyquist scheme work well. With two snapshots (K=2), the super-Nyquist estimate is bad. However, with K=4 and K=10 the estimate stabilizes. Therefore, low latency estimation is possible. Now, let us add another frequency band at 300Hz. The results are shown in Fig.6. Observe that the prototype co-prime scheme fails due to aliasing while the super-Nyquist scheme is still capable of estimating the three bands. This is because for the prototype co-prime scheme [$f_s = 500$Hz, $f_{max} = 250$Hz] while for super-Nyquist co-prime [$f_{ss} = 1000$Hz, $f_{max_{ss}} = 500$Hz]. This is achieved with the same prototype co-prime samplers. However, here the second sub-sampler has an analog delay of 0.5d when compared with the prototype co-prime sampler. It may be noted that the computational complexity for autocorrelation estimation is similar to that derived for the non-blind system [14][21].

So, will any value of (M, N) work? To answer this question, few more examples are considered in Fig. 7. (3, 4) seems to be good but (3, 5) and (5, 3) give bad results. Furthermore, (4, 3) in Fig. 6 is better than (3, 4). Therefore, appropriate values of (M, N) will have to be investigated in the future for the super-Nyquist scheme. Relative amplitude of main lobe and side-lobe peaks of the bias window can be used as a criterion as described in [14].

We end this section with another example of spectral estimation with bands at [0.1, 0.3, 0.6, 0.9]. The results are shown in Fig. 8. This again demonstrates the superiority of the proposed super-Nyquist co-prime sampling. It is important to note that the super-Nyquist as well as the prototype co-prime power spectrum is approximated by the convolution of the true (Nyquist) spectrum with the bias window. Therefore, the location of the peaks can affect the results.

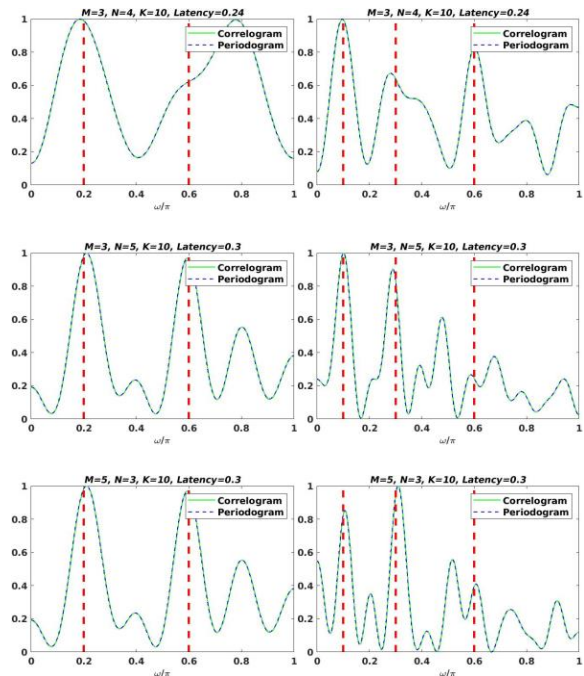

Fig. 7: Spectral estimation using super-Nyquist and prototype co-prime samplers: Three peaks, several (M, N) pairs.

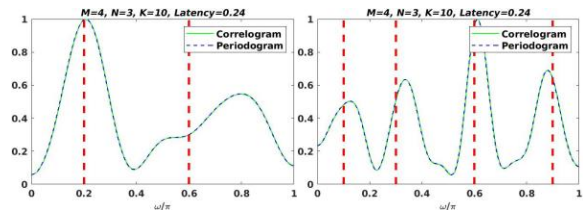

Fig. 8: Spectral estimation using super-Nyquist and prototype co-prime samplers: Four peaks, (M=4, N=3).

## 6. SUPER-NYQUIST SENSING WITH MULTIPLE PERIODS

The theory of prototype co-prime arrays with multiple periods was developed in [14]. The difference set analysis, expressions for weight function, correlogram bias window, and covariance were developed. Along similar lines, the concept of super-Nyquist co-prime sampling with multiple periods is shown in Fig. 9(a).

The expression for the weight function and bias window is given by (6) and (7). It can be derived similar to that in Appendix A. The bias window for periods r =1 to 4 is shown in Fig. 10 for the super-Nyquist co-prime scheme with multiple periods. The main lobe width and peak side-lobe reduces. Large improvement is observed for lower values of r. The

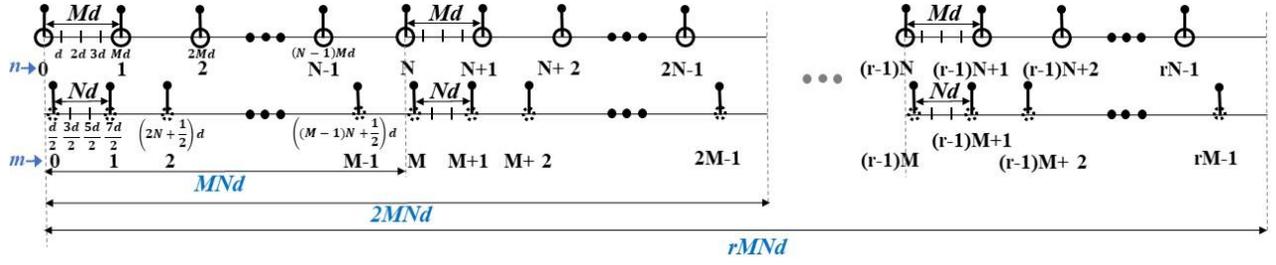
(a) Super-Nyquist co-prime sampling with multiple periods

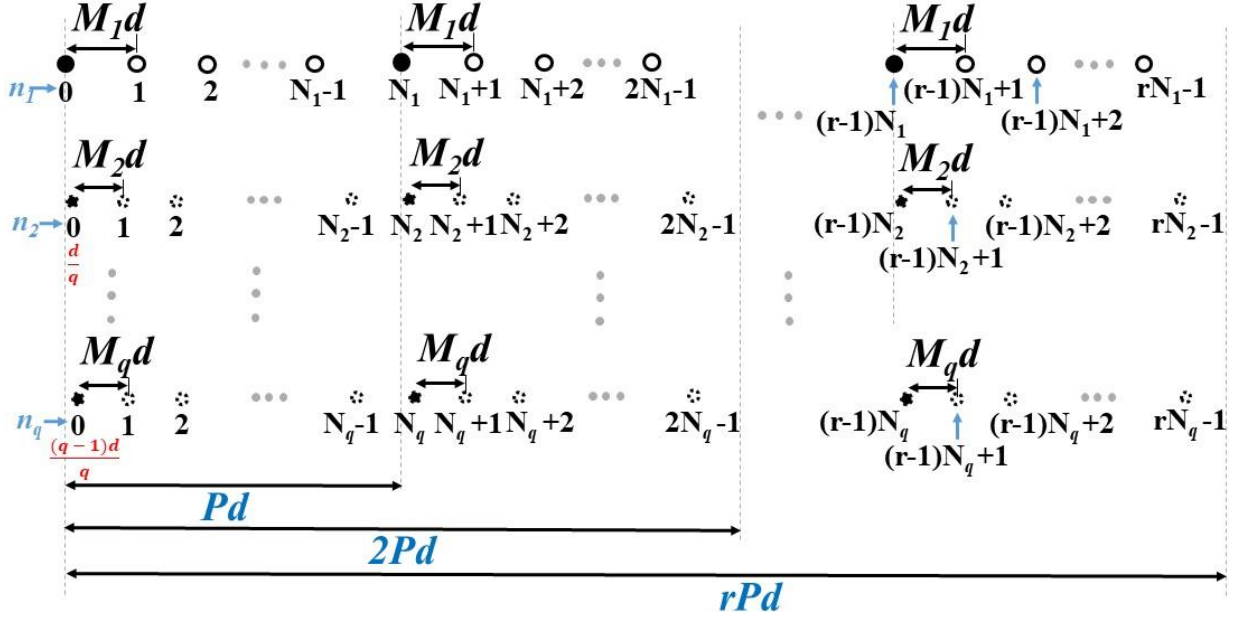
(b) Super-Nyquist n-tuple or multi-level prime sampling with multiple periods

Fig. 9: Super-Nyquist co-prime samplers with multiple periods and multiple levels.

trend is similar to that of the prototype co-prime samplers with multiple periods, which has been studied in the literature. As noted earlier, the bias window should resemble an impulse function as far as possible. The use of multiple periods seems to be a step in the right direction. The computational complexity would be similar to the non-blind system for multiple-period based co-prime sensing [29].

$$z(l) = \sum_{n=-rN+1}^{rN-1}(rN-|n|)\delta(l-2Mn)$$
$$+ \sum_{m=-rM+1}^{rM-1}(rM-|m|)\delta(l-2Nm)$$
$$+ \sum_{n=0}^{rN-1}\sum_{m=0}^{rM-1}\delta(|l|-|2Mn-2Nm-1|)$$

(6)

$$W(e^{j\omega}) = \frac{1}{s}\left\{\left|\frac{\sin \omega rMN}{\sin \omega M}\right|^2 + \left|\frac{\sin \omega rMN}{\sin \omega N}\right|^2 + 2\cos(\omega(M-N+1))\frac{(\sin \omega rMN)^2}{\sin \omega M \sin \omega N}\right\}$$

(7)

Before we conclude, let us have a look at another interesting structure. It is called the n-tuple or multi-level prime arrays/samplers [27][28] for one period. Super-Nyquist n-tuple/multi-level prime sampling is proposed in Fig. 9(b) with r =1. It can also be extended to multiple periods (r >1) as shown. Note that multi-level implies multiple (greater than two) sub-samplers (or sub-arrays). "n-tuple" in the original paper refers to 'n' sub-arrays. Here, 'q' denotes number of sub-samplers. The Fig. 9(b) shows q sub-samplers, with relatively prime pairs $\{N_i\}$ where $1 \leq i \leq q$. The inter-element spacing between two samples of the $i$th sub-sampler is $M_i = \prod_{\substack{k=1 \\ k \neq i}}^{q} N_k$. 'P' represents the multi-level period i.e. product of all $\{N_i\}$'s, given by $P = \prod_{k=1}^{q} N_k$.

Note that the second sub-sampler is placed at an offset of $\frac{d}{q}$, the third at $\frac{2d}{q}$, and the qth sub-sampler at $\frac{(q-1)d}{q}$. Therefore, super-Nyquist sampling period is $d_{ss} = \frac{d}{q}$ and hence, $f_{ss} = qf_s$. This implies that if the co-prime scheme could estimate the statistics of a signal with bandwidth B, then the super-Nyquist n-tuple structure can estimate a signal with bandwidth qB. A mathematical analysis of this structure along with simulations may be developed in the future.

From the authors limited knowledge, the difference set, expressions of weight and bias function for the standard n-tuple or multi-level sampler has not been developed in the literature. Once the theory is developed it could be easily extended to the super-Nyquist n-tuple/multi-level prime samplers.

## 7. CONCLUSION

This paper has proposed a novel super-Nyquist co-prime structure. The theory has also been considered. Few additional side-lobes are observed in the bias plots for the super-Nyquist scheme. Simulation results indicate good spectral estimation despite side-lobes. However, bias mitigation strategies can be explored in the future.

The same rate co-prime samplers (or multi-level samplers) is used but the proposed scheme has the ability to acquire the signals with higher bandwidths. However, the hardware implementation will require an analog delay of 0.5d between the two uniform co-prime samplers. This delay, in general is $\frac{d}{q}$, when q sub-samplers are employed. It is also necessary to maintain the synchronism which might be a challenge from an implementation point-of-view.

Here, the simulation results demonstrate the concept of spectrum sensing. Specific applications can be explored in the future. For example, in cognitive radios, secondary users need to scan a wide spectrum to detect holes for communication. In the future, researchers can examine the super-Nyquist co-prime structures using other spectral estimation strategies. For example, multiplicative processing, min processing, DFT filter banks, etc. [7][14][30][31]. Researchers may also wish to investigate subspace based spectral estimation techniques. However, the super-Nyquist structure described does not have a continuous difference range. The missing autocorrelation values may cause a problem for matrix inversion. Interpolation techniques noted in the paper may be employed to mitigate this problem.

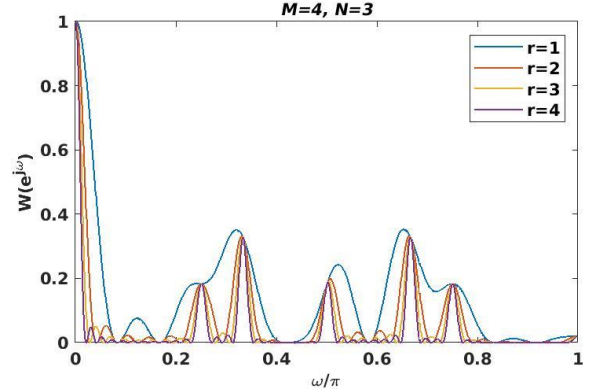

Fig. 10: Bias window for super-Nyquist co-prime samplers with multiple periods as in Fig. 9(a).

## APPENDIX A

### PROOF OF PROPOSITION I

Let A, B and C represent the first, second and third terms in equation (4). Let $\mathfrak{F}\{\cdot\}$ denote the Fourier transform, then:

$$\mathfrak{F}\{A\} = \sum_{l=-(L-1)}^{L-1} \sum_{n=-N+1}^{N-1} (N - |n|)\delta(l - 2Mn)\, e^{j\omega l}$$

$$= \sum_{n_1=0}^{N-1} \sum_{n_2=0}^{N-1} e^{j\omega 2M(n_1 - n_2)}$$

$$= \left(\frac{\sin(\omega MN)}{\sin(\omega M)}\right)^2$$

Similarly,

$$\mathfrak{F}\{B\} = \left(\frac{\sin(\omega MN)}{\sin(\omega N)}\right)^2$$

Lastly,

$$\mathfrak{F}\{C\} = \sum_{l=-(L-1)}^{L-1} \sum_{n=0}^{N-1} \sum_{m=0}^{M-1} \delta(|l| - |2Mn - 2Nm - 1|)\, e^{-j\omega l}$$

$$= \sum_{n=0}^{N-1}\sum_{m=0}^{M-1} e^{-j\omega(2Mn-2Nm-1)} + e^{j\omega(2Mn-2Nm-1)}$$

$$= \left(\frac{e^{j\omega}e^{j\omega M}}{e^{j\omega N}} + \frac{e^{-j\omega}e^{j\omega N}}{e^{j\omega M}}\right)\frac{\sin^2(\omega MN)}{\sin(\omega M)\sin(\omega N)}$$

$$= 2\cos\omega(M-N+1)\frac{\sin^2(\omega MN)}{\sin(\omega M)\sin(\omega N)}$$

By combining the above expressions, we get (4).